\documentclass[12pt]{article}
\usepackage{graphics,epsfig}

\def\dalemb#1#2{{\vbox{\hrule height .#2pt
        \hbox{\vrule width.#2pt height#1pt \kern#1pt
                \vrule width.#2pt}
        \hrule height.#2pt}}}

\let\a=\alpha \let\b=\beta \let\g=\gamma \let\d=\delta \let\e=\epsilon
\let\z=\zeta  \let\th=\theta  \let\k=\kappa
\let\l=\lambda \let\m=\mu  \let\x=\xi \let\p=\pi 
\let\s=\sigma \let\t=\tau   \let\c=\chi 
 \let\vep=\varepsilon
\let\w=\omega       \let\D=\Delta \let\Th=\Theta \let\L=\Lambda
\let\X=\Xi \let\P=\Pi \let\S=\Sigma  \let\Y=\Psi
\let\C=\Chi \let\W=\Omega
\let\la=\label \let\ci=\cite 
  
\def\nn{\nonumber} \def\bd{\begin{document}} \def\ed{\end{document}}
\def\ds{\documentstyle} \let\fr=\frac \let\bl=\bigl \let\br=\bigr
\let\Br=\Bigr \let\Bl=\Bigl
\let\bm=\bibitem
\let\na=\nabla
\def\tU{{\widetilde U}}
\let\pa=\partial \let\ov=\overline
\def\ie{{\it i.e.\ }}
\newcommand{\be}{\begin{equation}}
\newcommand{\ee}{\end{equation}}
\def\ba{\begin{array}}
\def\ea{\end{array}}
\def\ft#1#2{{\textstyle{{\scriptstyle #1}\over {\scriptstyle #2}}}}
\def\fft#1#2{{#1 \over #2}}
\def\F#1#2{{ F_{#1}^{(#2)} }}
\def\cF#1#2{{ {\cal F}_{#1}^{(#2)} }}

\def\={\, =\, }
\def\+{\, +\, }
\def\-{\, -\, }

\def\R{{\bf R}}
\def\sst#1{{\scriptscriptstyle #1}}
\def\oneone{\rlap 1\mkern4mu{\rm l}}
\def\e7{E_{7(+7)}}
\def\td{\tilde}
\def\wtd{\widetilde}
\def\im{{\rm i}}
\newcommand{\ho}[1]{$\, ^{#1}$}
\newcommand{\hoch}[1]{$\, ^{#1}$}
\newcommand{\bea}{\begin{eqnarray}}
\newcommand{\eea}{\end{eqnarray}}
\newcommand{\ra}{\rightarrow}
\newcommand{\lra}{\longrightarrow}
\newcommand{\Lra}{\Leftrightarrow}
\newcommand{\ap}{\alpha^\prime}
\newcommand{\bp}{\tilde \beta^\prime}
\newcommand{\cB}{{\cal B}}
\newcommand{\cO}{{\cal O}}
\newcommand{\vecx}{\vec{x}}
\newcommand{\vecy}{\vec{y}}
\newcommand{\vecp}{\vec{p}}
\newcommand{\vecq}{\vec{q}}
\newcommand{\tr}{{\rm tr} }
\newcommand{\Tr}{{\rm Tr} }

\newcommand{\cL}{{\cal L}}
\newcommand{\cA}{{\cal A}}
\newcommand{\cD}{{\cal D}}
\def\sst#1{{\scriptscriptstyle #1}}
\def\0{{\sst{(0)}}}
\def\1{{\sst{(1)}}}
\def\2{{\sst{(2)}}}
\def\3{{\sst{(3)}}}
\def\4{{\sst{(4)}}}
\def\5{{\sst{(5)}}}
\def\6{{\sst{(6)}}}
\def\7{{\sst{(7)}}}
\def\8{{\sst{(8)}}}
\def\ve{\varepsilon}
\def\vf{\varphi}
\def\F{\Phi}
\def\wg{\wedge}

\def \foot {\footnote}
\def \bi{\bibitem}

\def \tr {{\rm tr}}
\def \ha {{1 \over 2}}
\def \td {\tilde}
\def \ci{\cite}
\def \N {{\mathcal N}}
\def \ww {\Omega}
\def \const {{\rm const}}
\def \ss {\sum_{i=1}^3 }
\def \t {\tau}
\def\S{{\mathcal S} }
\def \nn {\nu}
\def \XX {{\rm X}}

\def \lra {\leftrightarrow}
\def \vom {{\bar \omega}}
\def \E {{\mathcal  E}} \def \J {{\mathcal  J}}
\def \YY {{\rm Y}}

\def \d {\del}
\def \rJ {{J}}
\def \sms {sigma models\ }
\def \sm {sigma model\ }
\def \L {\Lambda}
\def \gl {\ell}
\def \tr {{\rm tr\ }}
\def\z{\zeta}
\def\zi{\zeta_1}
\def\zii{\zeta_2}
\def\K{\mbox{K}}
\def\eE{\mbox{E}}   \def \vt {\vartheta}
\def \vr {\varrho}
\def \wup {w}

\def\dg{\dagger}
\def\a{\alpha}
\def\b{\beta}
\def\e{\varepsilon}
\def\p{\phi}
\def\ap{\alpha^\prime}
\def\I{{\cal I}}

\def\R{{\bf R}}
\def\Z{{\bf Z}}
\def\C{{\bf C}}
\def\P{{\bf P}}
\def\xb{{\bar X}}
\def\Tr{{\rm  Tr}}
\def\tr{{\rm  tr}}

\def \del{\partial}
\def \a {\alpha}
\def \aa {{\a'}}
\def\g{\gamma}
\def\s{\sigma}
\def\z{\zeta}
\def\zi{\zeta_1}
\def\zii{\zeta_2}
\def\ov{\over}

\def\I{{\cal I}}
\def\J{{\mathcal J}}
\def \ok {{1\ov \k}}
\def\LL{{\mathcal L }}
\def \jL {{J}}
\def \om {\omega}
\def \cL {{\mathcal L}} \def \cH {{\mathcal H}}
\def\E{{\mathcal E}}
\def\w{\omega}
\def\b{\beta}
\def\l{\lambda}
\def\eps{\epsilon}
\def\vep{\varepsilon}
\def \De {{\mathcal D}}

\def  \Jt {  {J}_{\rm tot}    }

\def \k {\kappa}
\def\foot{\footnote}
\def \four{{\textstyle {1\ov 4}}}
 \def \third { \textstyle {1\ov 3
}}
\def\det{\hbox{det}}
\def \ci {\cite}

\def \foot {\footnote}
\def \bi{\bibitem}

\def \tr {{\rm tr}}
\def \ha {{1 \over 2}}
\def \tid {\tilde}
\def \vv {{\rm v}}
\def \tl {{\tilde \l}}
\def \XX {{\rm X}}
\def \ta {{\tilde \a}}
\def \fo { {1\ov 4}}
\def \ep {\epsilon}
\def \inti {{\int^{2\pi}_0 {d \sigma \ov 2 \pi}}}

\def \d {\partial}
\def \K {{\rm S}}
\def \el {\ell}
\def \Tr {{\rm Tr}}
\def \P {\Phi}
\def \l  {\lambda}
\def \tl {{\tilde \l}}
\def \bl {{\tilde \l}}
\def \const {{\rm const}}
\def \V {v}

\def \bv {v^*}
\def \vv {{\rm v}}
\def \LL {{\mathcal L}}
\newcommand{\PV}[1]{P_{\!\!_{V_{#1}}}}

\def \bL {\ell}
\def \M {{\mathcal M}}
\def \N {{\mathcal N}}
\def \S {{\rm S}}
\def \vn {\vec n}
\def \tl {\td \l}
\def \td {\tilde}
\def \Prod {\Pi}
\def \O {{\mathcal O}}
\def \Q {{\rm  Q}}
\def \D {\Delta}
\def \N {{\mathcal N}}
\def\tN{{\tilde N}}

\def \m {\mu}
\def \vs {\vec \s}
\def \ie {i.e.}

\def \cD {{\cal D}}

\def  \le  {\l_{\rm eff}}

\def \rS {{\rm S}}
\def\as{{\a}}
\newcommand{\bra}[1]{\mbox{$\langle #1 |$}}
\newcommand{\ket}[1]{\mbox{$| #1 \rangle$}}

\newcommand{\auth}{AUTHORS}

\def\thb{\bar{\theta}}
\def\Thb{\bar{\Theta}}
\def\barp{\bar{p}}
\def\barq{\bar{q}}
\def\barc{\bar{c}}
\def\bard{\bar{d}}
\def\e{\epsilon}

\def \bi{\bibitem}
\def \la {\label}

\def \l {\lambda}
\def\foot{\footnote}
\def \tl  {{\tilde \l}}
\def \sql {{\sqrt \l}}
\def \adss {$AdS_5 \times S^5$\ }
\newcommand{\rf}[1]{(\ref{#1})}
\def \ov {\over}

\def\th{\theta}
\def\Th{\Theta}
\def\vth{\vartheta}
\def\btheta{{\bar\theta}}
\def\ttheta{{{\tilde\theta}}}
\def\bttheta{{{\bar\ttheta}}}
\def\vth{\vartheta}

\def\ra{\rightarrow}
\def\N{{\cal N}}
\def\F{{\cal F}}
\def\uM{\underline{M}}
\def\uN{\underline{N}}
\def\uP{\underline{P}}
\def\cc{\circ}
\def\eqv{\equiv}

\def\ni{\noindent}
\def \ha{{1\ov 2}}
\def \bw {{\rm w}}

\def\r{{\rm r}}

\def\a{{\rm\bf a}}
\def\b{{\rm\bf b}}
\def\c{{\rm\bf c}}

\def\Y{{\rm Y}}
\def\X{{\rm X}}
\def\tY{\tilde{\rm Y}}
\def\tX{\tilde{\rm X}}
\def\dY{\dot{\rm Y}}
\def\dX{\dot{\rm X}}

\def \J {\mathcal{J}}
\def \del {\partial}

\def\dF{\dot{F}}
\def\dG{\dot{G}}
\def\df{\dot{f}}
\def \E {{\cal E}}

\def \S {{\cal S}}
\def \J {{\cal J}}

\def\ms{\mathcal{S}}
\def\mj{\mathcal{J}}
\def\soj{\fr{\ms}{\mj}}
\def \R {{\bf R}}
\def \om {\omega}
\def \tH {\widetilde H}
\def \bE {\bar E}
\def \x {{\cal X}}

 \def \bb {\bar \beta}
\def \W {{\cal E}}

\def \bi{\bibitem}
\def \la {\label}

\def \l {\lambda}
\def\foot{\footnote}
\def \tl  {{\tilde \l}}
\def \sql {{\sqrt \l}}
\def \sqtl {{\sqrt {\tilde \l}}}

\def \HH {{\rm E}}

\def \adss {$AdS_5 \times S^5$\ }

\def \D {\Delta}
\def \thet {\theta}
 \def \t {\tau}
 \def \p {\phi}
 \def \r {\rho}
 \def \rN {{\rm N}}
 \def\tw{{\tilde w}}
 \def\hJ{{J}}
 \def\hw{{w}}
 \def\hl{{\lambda}}
 \def\hth{{\theta}}
 \def\NN{{\cal N}}
 \def \bv {{ \bar w}}
\def \vn {{\vec n}}

\def \ov {\over}

\def \varpi {{\rm w}}
\def \OO {{\cal O}}

\begin{document}
\overfullrule=0pt
\parskip=2pt
\parindent=12pt
\headheight=0in \headsep=0in \topmargin=0in \oddsidemargin=0in

\vspace{ -3cm} \thispagestyle{empty} \vspace{-1cm}

\begin{flushright}

\end{flushright}

\begin{center}

{\Large\bf Quantum  Landau-Lifshitz model at four loops: $1/J$ and
$1/J^2$ corrections to BMN  energies

 \vspace{0.01cm}
 }

 \vspace{.5cm} {
 A. Tirziu$^{}$\footnote{tirziu@mps.ohio-state.edu}
}\\
 \vskip 0.3cm

{\em Department of Physics, The Ohio State University,\\
Columbus, OH 43210, USA\\
}

\end{center}

 \begin{abstract}
In a previous paper (hep-th/0510080) the effective Landau-Lifshitz
(LL) Lagrangians in the $SU(2)$ sector coming from string theory
and gauge theory have been found to three loops in the effective
expansion parameter $\tilde\lambda=\lambda/J^2$. In this paper we
continue this study and find the effective Landau-Lifshitz
Lagrangians to four loops. We extend to four-loops
$\tilde\lambda^4$ the computations of $1/J$ and $1/J^2$
corrections to BMN energies done in hep-th/0510080 to three-loops.
We compare these corrections obtained from quantum
``gauge-theory'' LL action with the corrections obtained from the
conjectured Bethe ansatz
 for the long range spin chain
 representing perturbative \mbox{large $N$} $\NN=4$ Super Yang-Mills
in the $SU(2)$ sector and find perfect matching to four loops
$\tilde\lambda^4$. We compare also the $1/J$ and $1/J^2$
corrections obtained from quantum ``string-theory''  LL action
with those obtained from  the ``quantum string" Bethe ansatz and
again find perfect matching to four-loops.

\end{abstract}
\newpage

\setcounter{equation}{0} \setcounter{footnote}{0}
\setcounter{section}{0}

\renewcommand{\theequation}{1.\arabic{equation}}
 \setcounter{equation}{0}

\section{Introduction}

AdS/CFT correspondence implies that the quantum energy of a string
state matches the quantum dimension of a corresponding operator in
the dual $\NN=4$ Super Yang-Mills theory. Under certain conditions
a class of string quantum states can be treated as semiclassical
solutions and their classical energy can be compared with the
dimension of corresponding operators on the gauge theory side.
Going beyond semiclassical approximation it turns out that quantum
corrections to semiclassical solutions of strings propagating in
$AdS_{5}\times S^{5}$ play an important role in the AdS/CFT
duality \ci{bmn,gkp,ft1}.

The ``three-loop discrepancy'' between gauge and string theory
predictions is by now a well known fact: it is present \cite{ss}
for semiclassical spinning string solutions \cite{ft2}, and also
in the computation of the $1/J$ correction to the two-impurity BMN
state \cite{callan}. The computations of $1/J$ corrections to BMN
in \cite{callan} and to circular string solution in \cite{ft2}
were done using full superstring theory, i.e. using the full set
of world sheet fields, both bosons and fermions. Attempts to
compute the next correction, i.e. $1/J^2$, to the BMN energies
proved to be difficult \cite{Swanson}, and have not  yet been
obtained from
 a full superstring computation. An alternative and much easier
approach to compute $1/J^{n}$ corrections to BMN energies have
been developed in \cite{mtt} for the $SU(2)$ sector. It was shown
in \cite{btz,mtt} that by using the effective Landau-Lifshitz (LL)
action one can ignore all fermions and bosonic modes outside the
$SU(2)$ sector. The only effect of these modes in the effective LL
theory in the $SU(2)$ sector is the proper regularization, which
one needs to consider. As shown in \cite{mtt} this can be
accomplished with a combination of normal ordering and
$\zeta$-function regularization. Let us point out that this
discussion is valid only for analytic terms in $\tl$. Non-analytic
terms, i.e non-integer powers in $\tl$ terms were found in
\cite{bt,nz} (see also \cite{nzz} for an analysis of the validity
of zeta-function regularization), and it was shown in \cite{mtt1}
that they cannot be obtained correctly in the LL approach
regardless of the regularization used. In this paper we will
consider only analytic terms.

On string side an LL action in the $SU(2)$ sector can be obtained
by considering fast moving strings propagating in the $S^{3}$
subspace of $S^{5}$ \cite{kru,krt,kt}. The resulting string LL
action have been used in \cite{mtt1} to compute the $1/J^2$
corrections to BMN energies to three-loops in $\tl^3$. Results
from the string LL action have been compared with results from the
string ``quantum" Bethe ansatz of Arutyunov, Frolov and Staudacher
(AFS) \cite{afs}, and for the $1/J^2$ corrections for $M$-impurity
  BMN states
   perfect agreement has been found. In this paper we continue
   this comparison to four loops $\tl^4$. However, in order to do
   this we need to find first the string LL at four-loops. We find
   it indirectly: first at 4-loops we construct the LL action
   by including all
   possible eight derivative terms; secondly we fix the
   coefficients so that the energy as computed from LL matches the string
   energy for circular solution, and also the $1/J$ and $1/J^2$
   corrections to BMN energies as obtained from quantum string LL
   match those obtained by string Bethe ansatz in
   \cite{afs,mtt1}. All these conditions are consistent and determine uniquely
   the string LL Lagrangian.

On gauge theory side the gauge LL action at one loop in the
$SU(2)$ sector is the effective action for the ferromagnetic
Heisenberg spin chain in the continuum limit, with higher
derivative counterterms to account for lattice effects. Going to
higher orders in $\tl$ corresponds in terms of spin chain to going
beyond nearest neighbor interactions. An all-loop Bethe ansatz
that  preserves BMN scaling to all loops in the thermodynamic
limit was proposed by
 Beisert, Dippel and Staudacher  (BDS)  \cite{bds}. In order to compare
 results between this BDS Bethe ansatz and an effective action
 calculation one needs to find extensions of gauge LL action to higher loops.
 The LL action from gauge theory was derived to two-loops in
\cite{krt} where it was shown to be the same as the string LL
action derived from string theory. In \cite{mtt1} the gauge LL
action to three-loops was found by indirect methods. In this paper
 we extend that method to four loops and proceed as for the string LL:
 we compare results from gauge LL action for circular
string, $1/J$ and $1/J^2$ corrections to BMN energies obtained
from quantum gauge LL to results obtained from BDS Bethe ansatz
for operators that are duals to circular strings. This allows us
to determine uniquely the gauge LL Lagrangian to four-loops. Also
we find perfect matching between the $1/J$ and $1/J^2$ corrections
to BMN energies as obtained from quantum gauge LL and BDS Bethe
ansatz.

The manifestation of the ``three-loop discrepancy'' at the
classical level of LL actions was obtained in \cite{mtt1} where
gauge and string LL actions were found to differ at three-loops
$\tl^3$. In this paper we find that the ``three-loop discrepancy''
continues at four-loops, as expected. The $1/J$ and $1/J^2$
corrections to BMN energies, which can be found from quantum LL,
also start differing at three-loops and continue to differ at
four-loops.

This paper is organized a follows: In Section 2 we describe the
gauge and string LL actions and find some of the unknown
coefficients by using the circular string solution. In Section 3
after reviewing the quantization of LL action developed in
\cite{mtt}, we compute the $1/J$ and $1/J^2$ corrections in
Section 3.1 and 3.2, respectively. In section 4 we collect the
results and find completely both the gauge and string LL actions.
In Appendix A we describe details of the computation of energy of
circular string by using LL action, string theory and BDS Bethe
ansatz. In Appendix B we present some details of the computation
of $1/J^2$ corrections to BMN energies from quantum LL.

\renewcommand{\theequation}{2.\arabic{equation}}
 \setcounter{equation}{0}

\section{Classical LL action to $\tl^4$ order}

The main steps as well as notations in this section and throughout
the paper are as in \cite{mtt1}. We start with the string LL
action in the $SU(2)$ sector \cite{mtt1} (for a review see also
\cite{tse2, mtt})
\begin{equation}\label{aal}
S=J\int dt \int_{0}^{2\pi}\frac{d\sigma}{2\pi}\ L \ ,
\end{equation}
 where
the  Lagrangian to four loops is
\begin{eqnarray} \label{4loop}
L&=&  \vec C (n)   \cdot    \del_t \vn -\ \frac{1}{4}\vn\
\bigg(\sqrt{1-\frac{\lambda}{\pi^2}\sinh^2 \frac{\pi
\partial_{1}}{J}}-1\bigg)
 \ \vn\ -
\frac{3\tilde{\lambda}^2}{128}(\partial_{1}\vn)^4\\
&-&\frac{\tilde{\lambda}^3}{64}\left[\ \a \ (\partial_{1}\vn)^2
(\partial_{1}^2 \vn )^2\ +\ \b\ (\partial_{1}\vn
\partial_{1}^2 \vn)^2\  + \  \c \
(\partial_{1}\vn)^6\right]\nonumber\\
&+&\tl^4 \bigg[ \a_{1} \ (\partial_{1}^2 \vec{n})^4  +\a_{2} \
(\partial_{1} \vec{n})^2(\partial_{1}^2 \vec{n} \partial_{1}^4
\vec{n}) + \a_{3} \
(\partial_{1}\vec{n}\partial_{1}^5\vec{n})(\partial_{1}\vec{n})^2\nonumber\\
&+& \a_{4} \ (\partial_{1}^3 \vec{n})^2(\partial_{1}\vec{n})^2  +
\ \a_{5} \ (\partial_{1}\vec{n})^4(\partial_{1}^2 \vec{n})^2+ \
\a_{6} \ (\partial_{1}\vec{n}\partial_{1}^2
\vec{n})^2(\partial_{1}\vec{n})^2+ \ \a_{7} \
(\partial_{1}\vec{n})^8\bigg] + O (\tl^5) \ . \nonumber
\end{eqnarray}
where at the three loops we introduced all structures with six
derivatives, and at four loops all structures with eight
derivatives up to integrations by parts. As in \cite{mtt1, mtt}
here $d C= \epsilon^{ijk} n_i d n_j \wedge d n_k$, i.e. $\vec C$
is a monopole potential on $S^2$, $\tl=\lambda/J^2$, where
$\lambda \sim 1/(\alpha')^2$, $\vn (t,\s)$ is a unit vector. At
quadratic order in $\vn$ we considered the exact kinetic operator
corresponding to the exact spin-chain dispersion relation
\cite{bds} $\omega(n)=\sqrt{1+\frac{\lambda}{\pi^2}\sin^2
\frac{\pi n}{J}}$ in momentum representation
($p=\frac{2\pi}{J}n$). Writing the kinetic term in terms of $\tl$,
$J$ and expanding at large $J$ one obtains
\begin{equation}
\frac{1}{4}\vn \bigg(\sqrt{1-\frac{\lambda}{\pi^2}\sinh^2
\frac{\pi\partial_{1}}{J}}-1\bigg) \ \vn=\ \frac{1}{4}  \vn
\left(\sqrt{1-\tilde{\lambda}\partial_{1}^2}-1\right)\vn -
\frac{\tilde{\lambda}\pi^2}{24J^2} \ \vn
\frac{1}{\sqrt{1-\tilde{\lambda}\partial_{1}^2}}\partial_{1}^4 \vn
+...\label{kinetic}
\end{equation}
The first term in this expansion can be also obtained from the
classical string action by integrating out all superstring
world-sheet fields except $\vn$. The second term is a quantum
correction\footnote{One can view the factor of $J$ in front of the
LL action as the inverse Planck constant. This becomes clearer
when we introduce fluctuations in the next section.}, suggesting
that it should come from considering the full superstring theory.
Therefore we expect that the exact kinetic term in (\ref{4loop})
appears in the quantum effective field theory derived from the
full superstring theory. Since the second term in (\ref{kinetic})
can only give contribution starting at $\tl/J^2$, we add this
contribution when computing this correction only, while
considering for simplicity the action (\ref{4loop}) with the
kinetic term given by the first term in (\ref{kinetic}).

The coefficients $\a$, $\b$, $\c$ were fixed in \cite{mtt1} for
both string and gauge LL. Let us recall the values of the
``3-loop'' coefficients in the string and
  gauge theory LL Lagrangians \cite{mtt1}
\begin{equation}\la{oi}
\a _{s}=-\frac{7}{4}\ , \quad \quad \b_{s}=-\frac{25}{2}\ , \quad
\quad \c_{s}=\frac{13}{16}\ ,
\end{equation}
\begin{equation}\la{opim}
\a _{g}=-\frac{7}{4}\ , \quad \quad \b_{g}=-\frac{23}{2}\ , \quad
\quad \c_{g}=\frac{3}{4}\ .
\end{equation}
In this paper we want to fix the
coefficients $\a_{i}$, $i=1...7$. To do this we consider again, as
in \cite{mtt1}, the circular solution with two unequal spins
($J_{1} \ne J_{2}$). For this solution we compare the energy at
four loops ($\tl^4$) obtained from the string LL with the energy
obtained from string theory at this order. We present the details
in Appendix A. We obtain the following string LL coefficients
\begin{equation}
\a_{5}^{(s)}=\frac{119}{4096}\ , \quad \quad
\a_{7}^{(s)}=-\frac{323}{32768}\ , \quad \quad
\a_{1}^{(s)}-\a_{2}^{(s)}+\a_{3}^{(s)}+\a_{4}^{(s)}=-\frac{59}{2048}
\label{s1}
\end{equation}

On the other hand we can compare the energy for the circular
solution obtained from LL to the corresponding solution on gauge
theory side by using gauge BDS Bethe ansatz. We obtain then the
gauge LL coefficients
\begin{equation}
\a_{5}^{(g)}=\frac{111}{4096}\ , \quad \quad
\a_{7}^{(g)}=-\frac{267}{32768}\ , \quad \quad
\a_{1}^{(g)}-\a_{2}^{(g)}+\a_{3}^{(g)}+\a_{4}^{(g)}=-\frac{59}{2048}
\label{g1}
\end{equation}

We see that in both gauge and string LL we need more equations
among the coefficients to find them. Following the method in
\cite{mtt1} we compare the $\tl^4/J$ and $\tl^4/J^2$ corrections
to BMN vacuum obtained from quantum LL to the corresponding
results obtained from gauge (BDS) and string (AFS) Bethe ansatze.

As in  \cite{mtt1,mtt},  let us  now rewrite  the LL Lagrangian
\rf{4loop} in terms of two independent fields. Solving the
constraint $|\vn|^2 =1$  as
 $n_3=\sqrt{1-n_{1}^2-n_{2}^2}$  we get the following  $SO(2)$
 invariant expression for the
 Lagrangian in terms of $n_1$ and $n_2$   ($a,b=1,2$; \  $n^2= n_a n_a$)
\begin{equation}
L=\  h^2(n)\
  \epsilon_{ab}\dot{n}_{a}n_{b} - H (n_1,n_2) \ , \ \ \ \ \ \ \ \ \ \
 h^2(n)=\frac{1-\sqrt{1-n^2}}{2n^2}={1 \ov 4 } + { 1 \ov 16} n^2 + ...\ .
 \label{laag}
 \ee
We omitt the complicated form of $H(n_{1},n_{2})$ for simplicity.
To simplify the quantization of the LL Lagrangian  near a
particular solution it is useful
 to put it  into the standard canonical form \ci{mtt}
 by doing the field
redefinition $n_a \to z_a $
\begin{equation}
z_{a}=h(n)\ n_{a} \ , \ \ \ \ \ \ \ \ \ \ n_{a}= 2\sqrt{1-z^2}\
z_{a} \ , \la{nz} \ee to obtain
\begin{equation}
L=\epsilon_{ab}\dot{z}_{a}z_{b}  -H(z_1,z_2) \ .  \label{vv}
\end{equation}
Having the Lagrangian in the standard form $L=p \dot q - H(p,q)$,
the quantization is straightforward: we  promote $z_a$ to
operators, impose the canonical commutation relation (cf.
\rf{aal})
\begin{equation}
[z_1(t,\s),z_2 (t,\s')]=   i J^{-1} \pi \delta (\s - \s')\,,
\label{cann}
\end{equation}
 and then decide how
to order the ``coordinate'' and ``momentum''  operators in
$H(z_1,z_2)$.

\renewcommand{\theequation}{3.\arabic{equation}}
 \setcounter{equation}{0}
\section{Quantization near  BPS vacuum:
corrections\\ to BMN spectrum from  LL  Hamiltonian}

Following \ci{mtt,mtt1} we use the  LL action to compute quantum
 $1/J$  and $1/J^2$ corrections to the
  BMN spectrum of fluctuations near
 the  BPS vacuum solution
\begin{equation}
\vec n=(0,0,1)\ , \ \ \
\end{equation}
representing the massless geodesic  in $R_t \times S^3$.
 The $1/J$
  corrections can be found
from the  Bethe ansatz on the spin chain \ci{mz,beisert} or from a
direct superstring quantization \ci{parn,callan}. As explained in
\ci{mtt}, the derivation  from the  LL action turns out to be much
simpler
 than the   string-theory derivation.
Here we  will extend the method of \ci{mtt,mtt1} to $\tl^4/J$
 and  $\tl^4/J^2$ orders.\foot{Unfortunately, the exact
 (all order in $\tl$)
 form of the $\vec n^4$, $\vec n^6$ and $\vec n^8$ terms in the LL  action is not known,
 preventing us  from computing the $1/J$ and $1/J^2$ corrections to all
 orders in $\tl$.}

Expanding  near this vacuum corresponds to expansion near $n_a=0$
in \rf{laag}
 or $z_a=0$ in
\rf{vv}. Observing that  the factor $J$ in front of  the LL action
\rf{aal}
 plays the role of the inverse Planck constant,
it is natural to rescale  $z_a$ as
\begin{equation}
z_{1}= \frac{1}{\sqrt{J}}\ f\ ,\ \ \ \  \quad z_{2}=
\frac{1}{\sqrt{J}}\ g\ , \la{fg}
\end{equation}
so that powers of $1/J$ will play the role of coupling constants
for the fluctuations in the  non-linear  LL Hamiltonian.
 Expanding the Hamiltonian in (\ref{laag}), (\ref{vv}) to  sixth  order in the fluctuation
 fields $f,g$  we get
\begin{equation}
S=\int dt \int_{0}^{2\pi}\frac{d\sigma}{2\pi}\ ( 2g \dot{f} - H )
\ , \    \ \ \ \ \ \ \ H = H_2 + H_4 +H_6 + ... \ , \la{qua} \ee
\be H_2 = f \ (\sqrt{1-\tilde{\lambda}\partial_{1}^2}-1) \ f\ + \
g \ (\sqrt{1-\tilde{\lambda}\partial_{1}^2}-1) \ g,  \label{quad}
\ee
\begin{eqnarray}
 H_4=&& { 1 \ov J} \bigg\{ (f^2+g^2) (\sqrt{1-\tilde
 {\lambda}\partial_{1}^2}-1)(f^2+g^2)-
  \ f(f^2+g^2)(\sqrt{1-\tilde{\lambda}
 \partial_{1}^2}-1)f\nonumber\\
 &&\ \  \ -\
  \ g(f^2+g^2)(\sqrt{1-\tilde{\lambda}\partial_{1}^2}-1)g
 \ +\frac{3\tilde{\lambda}^2}
 {8}(f'^2+g'^2)^2 \label{quartic}
\\
 &&\ \ \ +\ \frac{\tilde{\lambda}^3}{4}\left[\b (f'f''+g'
 g'')^2+\a (f'^2+g'^2)(f''^2+g''^2)\right]-16 \tl^4 \bigg[\a_{1}(f''^2+g''^2)^2\nonumber\\
 &&\ \ \ +\a_{2}(f'^2+g'^2)(f'' f''''+g'' g'''') + \ \a_{3}(f'^2+g'^2)(f' f^{(5)}+g'
g^{(5)})\nonumber\\
&&\ \ \ + \ \a_{4}(f'^2+g'^2)(f'''^2+g'''^2)\bigg]\bigg\}
  +  O ( {\tl^5\ov J})
\ , \nonumber
\end{eqnarray}
\begin{eqnarray}
 H_6= &&{ 1 \ov J^2} \bigg\{ \ \  \frac{1}{4}f(f^2+g^2)
 (\sqrt{1-\tilde{\lambda}\partial_{1}^2}-1)[f(f^2+g^2)]  \nonumber\\
&&\ \ \ \ \ \ +\   \frac{1}{4}g(f^2+g^2)
 (\sqrt{1-\tilde{\lambda}\partial_{1}^2}-1)[g(f^2+g^2)]
 \nonumber\\
 &&\ \ \ \ \ \ -\ \frac{1}{4}f(f^2+g^2)^2 (\sqrt{1-\tilde{\lambda}\partial_{1}^2}-1)f
 -\frac{1}{4}g(f^2+g^2)^2
 (\sqrt{1-\tilde{\lambda}\partial_{1}^2}-1)g\nonumber\\
 &&\ \ \ \ \ \ +\ \frac{3\tilde{\lambda}^2}{4}\bigg[2(f'^2+g'^2)(f f'+g
 g')^2-(f^2+g^2)(f'^2+g'^2)^2\bigg]\nonumber\\
 &&\ \ \ \ \ \ +\ \frac{\tilde{\lambda}^3}{2}\bigg[2 \c (f'^2+g'^2)^3- \b
 (f'f''+g' g'')\bigg((f^2+g^2)(f' f''+g' g'')\nonumber\\
 &&\ \ \ \ \ \ - \  (f'^2+g'^2)(f f'+g g')-2(f f'+g g')(f f''+g
 g'')\bigg)\nonumber\\
 &&\ \ \ \ \  \ \ + \ \a (f'^2+g'^2)\bigg(2(f'^2+
 g'^2)^2+(f f''+g g'')^2+3 (f'^2+g'^2)(f f''+g g'')\nonumber\\
&& \ \ \ \    \ \ - \ 2(f' f''+g' g'')(f f'+g
g')-(f^2+g^2)(f''^2+g''^2)\bigg)\nonumber\\
&&\ \ \ \ \ \ + \ \a (f f'+g g')^2 (f''^2+g''^2) \bigg]- 32 \
\tl^4 \a_{1} \ (f''^2+g''^2)\bigg[2(f f''+g
g'')^2\nonumber\\
&& \ \ \ \ + \ \ 4(f'^2+g'^2)^2-4 (f' f''+g' g'') (f f'+g
g')+6(f'^2+g'^2)(f f''+g
g'')\nonumber\\
&& \ \ \ \ - \ \ (f^2+g^2)(f''^2+g''^2)\bigg]\nonumber\\
&& \ \ \ \ - \ \ 16 \ \tl^4  \a_{2}(f'^2+g'^2)\bigg[3(f f''+g
g'')(f''^2+g''^2)+12 (f f''+g g'')(f' f'''+g' g''')\nonumber\\
&& \ \ \ \ + \ \ 2(f f''+g g'')(f f''''+g
g'''')+6(f'^2+g'^2)(f''^2+g''^2)+16 (f'^2+g'^2)(f' f'''+g'
g''')\nonumber\\
&& \ \ \ \ + \ \ 3 (f'^2+g'^2)(f f''''+g g'''')-2(f^2+g^2)(f''
f''''+g'' g'''')\nonumber\\
&& \ \ \ \ - \ \ 4 (f'' f'''+g'' g''')(f f'+g g')-12 (f' f''+g'
g'')^2-4 (f' f''+g' g'')(f f'''+g g''')\nonumber\\
&& \ \ \ \ - \ \ 2 (f' f''''+g' g'''')(f f'+g g')\bigg]-32 \ \tl^4
\a_{2}(f f'+g g')^2(f'' f''''+g'' g'''')
\nonumber\\
&& \ \ \ \ + \ \ 16 \ \tl^4 \a_{3}(f'^2+g'^2)\bigg[2(f^2+g^2) (f'
f^{(5)}+g'
g^{(5)})\nonumber\\
&& \ \ \ \ - \ \ 10 (f' f''''+g' g'''')(f f'+g g')+10 (f' f'''+g'
g''')(f f''+g g'')\nonumber\\
&& \ \ \ \ + \ \ 30 (f' f''+g' g'')^2+10 (f' f''+g'
g'')(f f'''+g g''')\nonumber\\
&& \ \ \ \ + \ \ 15 (f'^2+g'^2)(f''^2+g''^2)+30 (f'^2+g'^2)(f'
f'''+g' g''')+5 (f'^2+g'^2)(f f''''+g g'''')\nonumber\\
&& \ \ \ \ - \ \ 30 (f f'+g g')(f'' f'''+g'' g''')-2 (f f'+g g')(f
f^{(5)}+g g^{(5)})\bigg]\nonumber\\
&& \ \ \ \ - \ \ 32 \ \tl^4 \a_{3}(f f'+g g')^2 (f'
f^{(5)}+g' g^{(5)})\nonumber\\
&& \ \ \ \ - \ \ 32 \ \tl^4 \a_{4} (f'^2+g'^2)\bigg[18(f' f''+g'
g'')^2+9
(f' f''+g' g'')(f f'''+g g''')\nonumber\\
&& \ \ \ \ + \ \ (f f'''+g g''')^2-3(f' f'''+g'
g''')(f'^2+g'^2)-3 (f' f'''+g' g''')(f f''+g g'')\nonumber\\
&& \ \ \ \ - \ \ 3 (f'' f'''+g'' g''')(f f'+g g') -(f^2+g^2)(f'''^2+g'''^2)\bigg]\nonumber\\
&& \ \ \ \ - \ \ 32 \ \tl^4 \a_{4} (f'''^2+g'''^2)(f f'+g g')^2-64
\ \tl^4
\a_{5}(f'^2+g'^2)^2 (f''^2+g''^2)\nonumber\\
&& \ \ \ \ - \ \ 64 \ \tl^4 \a_{6}(f'^2+g'^2)(f' f''+g' g'')^2
\bigg\} + O ( {\tl^5\ov J^2})
   \ .    \label{six}
\end{eqnarray}
Let us first consider the quadratic approximation. The linearized
equations of motion for the fluctuations are \be
\dot{f}=-(1-\sqrt{1-\tilde{\lambda}\partial_{1}^2})\ g\ , \quad\ \
\ \ \ \ \ \dot{g}=(1-\sqrt{1-\tilde{\lambda}\partial_{1}^2})\ f \
, \label{eqmotion} \ee and their  solution may be written as
\bea\label{fsol}
f(t,\sigma)&=&\frac{1}{2}\sum_{n=-\infty}^{\infty}(a_{n}e^{-i\omega_{n}t+in\sigma}+
a_{n}^{\dagger}e^{i\omega_{n}t-in\sigma})\ , \ \ \ \ \ \
\omega_{n} =  \sqrt{1+\tilde{\lambda}n^2}-1  \ ,\la{lo}\\
\label{gsol}
g(t,\sigma)&=&\frac{1}{2}\sum_{n=-\infty}^{\infty}(-ia_{n}e^{-i\omega_{n}t+in\sigma}+i
a_{n}^{\dagger}e^{i\omega_{n}t-in\sigma})\ , \la{loo} \eea for
real $f$ and $g$. Upon quantization  \rf{eqmotion} becomes the
equations of motion for the operators $f,g$
\begin{equation}\la{op}
\dot{f}=i[\bar H_2,f], \quad \quad \dot{g}=i[\bar H_2,g] \ , \ \ \
\ \ \ \bar H_2 \equiv   \int_{0}^{2\pi} \frac{d\sigma}{2\pi}\  H_2
\ ,
\end{equation}
provided we use the  canonical commutation relations in \rf{cann}
\begin{equation}\la{opi}
[f(t,\sigma),f(t,\sigma')]=0\  , \quad
[g(t,\sigma),g(t,\sigma')]=0\ , \ \ \ \
[f(t,\sigma),g(t,\sigma')]=i\pi \delta(\sigma-\sigma')\ .
\end{equation}
Then  the coefficients  in \rf{lo},\rf{loo}   satisfy
\begin{equation}\la{ui}
[a_{n},a_{m}^{\dagger}]=\delta_{n-m} \ ,
\end{equation}
so that   $a_{n}$ and $a_{n}^{\dagger}$ can be interpreted as
annihilation and creation operators, with the  vacuum state
$|0\rangle$ defined  by $a_n |0\rangle=0$, for all integer $n$.
 A general oscillator state is
\begin{equation}
|\Psi\rangle=\prod_{n=-\infty}^{\infty}
\frac{(a_{n}^{\dagger})^{k_{n}}}{\sqrt{k_{n}!}}|0\rangle \ .
\label{states}
\end{equation}
The  integrated  Hamiltonian $\bar H_2$  then becomes
\begin{equation}\label{ham2}
\bar H_2 = \sum_{n=-\infty}^{\infty}\ \omega_{n}\ a_{n}^{\dagger}
a_{n} \ , \ee where we have used the normal ordering to ensure
that the vacuum energy is   zero,  since the BMN vacuum is  a BPS
state in both gauge theory and string theory.

One also needs to impose the  extra constraint that the total
$\sigma$-momentum is zero  \cite{mtt}. For  physical oscillator
states we get \be \la{cons} \sum_{n=-\infty}^{\infty}n
a_{n}^{\dagger}a_{n} |\Psi\rangle =0\ , \ \ \ \ \ \ \ \
\sum_{n=-\infty}^{\infty}n k_n =0 \ . \ee Below we shall  consider
the
 ``$M$-impurity'' states as oscillator states with $k_n=1$:
 \begin{equation}
|M\rangle = a_{n_{1}}^{\dagger}...a_{n_{M}}^{\dagger}|0\rangle \ ,
\end{equation}
 where for  simplicity we  shall assume that  all $n_j$  are
 different (generalization to  states
 with several equal $n_j$ is straightforward,   at
 least for  $1/J$ corrections).
Then the  zero-momentum condition \rf{cons}   gives
\begin{equation}\label{momcond}
\sum_{j=1}^{M}n_{j}=0 \ ,
\end{equation}
and the  leading  term in the energy of an
 $M$-impurity state takes  the familiar  form  \ci{mets,bmn}
\begin{equation}
E^{(0)} = \langle  M|\bar H_2 |M\rangle=\sum_{j=1}^{M}(\sqrt
{1+\tilde{\lambda}n_{j}^2}-1)\ . \la{enn}
\end{equation}

Let us also  compute the  difference  of spins and obtain
\cite{mtt}
\begin{equation}
J_{1}-J_{2}= J-2 \sum_{n=-\infty}^{\infty}a_{n}^{\dagger}a_{n}\ .
\end{equation}
Applied to $M$-impurity state
 the above
relation gives $J_1-J_2=J-2M$. Since  $J_{1}+J_{2}=J$ we have
\begin{equation}
J_1=J-M\ ,\ \ \ \ \ \  \quad J_{2}=M \ .
\end{equation}
In the LL  approach we use $J= J_1 + J_2 $ as a natural total
angular momentum, corresponding to a ``fast'' collective
coordinate. $M$ is  a characteristic of a particular state, while
  $J$ enters into the background-independent form  of the LL
action  \rf{aal}.
 This is
in line  with gauge/spin chain intuition,  where the use  of total
$J$ or spin chain length as the   state-independent parameter is
natural. The corresponding gauge-theory states are Tr$(\Phi_1
^{J_{1}}\Phi_2 ^{J_{2}})+ ...$, and $J$ plays the role  of the
length of the spin chain and $M$ is the number of magnons.

\subsection{$1/J$ corrections to the BMN spectrum}

In this section we fix more unknown coefficients by comparing
$1/J$ correction to BMN vacuum obtained from the quantum string
LL, to the corresponding correction at $\tl^4$ obtained in
\cite{afs} from string Bethe ansatz. For the gauge LL coefficients
we compare with $1/J$ corrections obtained from gauge Bethe
ansatz. The $\tl^4/J$ correction obtained in \cite{afs} is
\begin{eqnarray}
\frac{\tl^4}{16 J}\bigg[3\sum_{k=1}^M n_{k}^3 \sum_{j=1}^M
n_{j}^5+\left(\sum_{j=1}^M n_{j}^4\right)^2+2 \sum_{k=1}^M n_{k}^6
\sum_{j=1}^M n_{j}^2-16 \sum_{k=1}^M n_{k}^8\bigg] \ .
\label{string1J}
\end{eqnarray}
Note that this matches the result obtained in \cite{swanson} from
string theory.

On the other hand the $1/J$ correction to BMN on the gauge theory
side at four loops ($\tl^4$) can be obtained from the BDS Bethe
ansatz \cite{bds}
\begin{eqnarray}
\frac{\tl^4}{16 J}\bigg[6\sum_{k=1}^M n_{k}^3 \sum_{j=1}^M
n_{j}^5-16 \sum_{k=1}^M n_{k}^8\bigg] \ . \label{gauge1J}
\end{eqnarray}

\bigskip

We compute now the $1/J$ correction to BMN vacuum by using the
quantum LL. Following the method developed in \cite{mtt} we add
the four loop contribution to the three loop one already computed
in \cite{mtt1}. We recall that at quartic order in fluctuations
one is to use normal ordering. We obtain then the following normal
ordered Hamiltonian \be \bar H_4= \frac{1}{J} \sum_{n,m} \
 h_{nm} \ a_{n}^{\dagger}a_{m}^{\dagger}a_{n}a_{m}\ , \la{hi} \ee
\bea h_{nm}= &&1+\sqrt{1+\tilde{\lambda}(n-m)^2}-
2\sqrt{1+\tilde{\lambda}n^2} \nonumber \\
&& \ \ +\  \frac{3\tilde{\lambda}^2}{4}n^2
m^2+\frac{\tilde{\lambda}^3}{16 } n^2 m^2 [  2\a ( n+ m)^2 +  \b
(n-m)^2  ]\nonumber\\
&& \ \  -\ 16 \tl^4 \bigg[n^4 m^4 \bigg(2 \a_{1}+\a_{4}\bigg)+n^6
m^2
\bigg(-\a_{2}+2 \ \a_{3}+\a_{4}\bigg)\nonumber\\
&& \ \ -\ n^5 m^3 \a_{2}\bigg] +O ({\tl^5}) \ . \la{hii} \eea
Computing the expectation value of $\bar{H}_{4}$ and expanding to
order $\tl^4$ we obtain
\begin{eqnarray}\la{ooo}
&&E^{(1)}= \langle  M|\bar H_4 |M\rangle= { 1 \ov J} \bigg\{ \tl
\sum_{i=1}^{M}n_i^2- \tilde{\lambda}^2  \sum_{i=1}^{M}n_i^4 +  { 1
\ov 8} ( 1 - 4 \a ) \tl^3 \sum_{i=1}^{M}n_i^6 \nonumber\\   && \ \
+ \ \frac{\tilde{\lambda}^3}{16}
 \sum_{i,j=1}^{M}  n_{i}^2n_{j}^2 \bigg[
   (2 \a + \b  + 15)  (n^2_i + n^2_j )  + 2 ( 2 \a - \b -10)
    n_i n_j \bigg]\nonumber\\
    && \ \ +16\tl^4 \bigg[-
    \bigg(2\a_{1}+\a_{4}+\frac{175}{1024}\bigg)\bigg(\sum_{i=1}^{M}n_i^4\bigg)^2\nonumber\\
    && \ \
    +\bigg(\a_{2}-2\a_{3}-\a_{4}-
    \frac{35}{256}\bigg)\sum_{i=1}^{M}n_i^6
    \sum_{k=1}^{M}n_k^2+\bigg(
    \a_{2}+\frac{35}{128}\bigg)\sum_{i=1}^{M}n_i^5
    \sum_{k=1}^{M}n_k^3\nonumber\\
    && \ \ +\bigg(2\a_{1}-2\a_{2}+2\a_{3}+2\a_{4}-
    \frac{5}{1024}\bigg)\sum_{j=1}^{M}n_j^8\bigg]
 +O(\tilde{\lambda}^5)   \bigg\} \ .
\end{eqnarray}
Comparing this expression at four loops with the result in
(\ref{string1J}) we obtain $\a_{2}^{(s)}$ and two independent
equations
\begin{equation}
\a_{2}^{(s)}=-\frac{67}{256} \label{s2}
\end{equation}
\begin{equation}
2 \a_{1}^{(s)}+\a_{4}^{(s)}=-\frac{179}{1024}\ , \quad \quad 2
\a_{3}^{(s)}+\a_{4}^{(s)}=-\frac{104}{256} \ . \label{s3}
\end{equation}
We see that combining the above equations we obtain the last
equation in (\ref{s1}), so that in fact we have only 2 independent
equations relating $\a_{1}^{(s)}$, $\a_{3}^{(s)}$ and
$\a_{4}^{(s)}$.

Comparing (\ref{ooo}) at four loops with the result in
(\ref{gauge1J}) we again obtain $\a_{2}^{(g)}$ and two independent
equations
\begin{equation}
\a_{2}^{(g)}=-\frac{64}{256} \label{g2}
\end{equation}
\begin{equation}
2 \a_{1}^{(g)}+\a_{4}^{(g)}=-\frac{175}{1024}\ , \quad \quad
2\a_{3}^{(g)}+\a_{4}^{(g)}=-\frac{99}{256} \ . \label{g3}
\end{equation}
We see that combining the above equations we obtain the last
equation in (\ref{g1}), so that, as in the string case, we have
only 2 independent equations relating $\a_{1}^{(g)}$,
$\a_{3}^{(g)}$ and $\a_{4}^{(g)}$.

In both string and gauge cases we need 2 more equations to
determine the still unknown coefficients $\a_{1}$, $\a_{3}$,
$\a_{4}$ and $\a_{6}$.

\subsection{$1/J^2$ corrections to the BMN spectrum}

To find $1/J^2$ corrections we follow the method in
\cite{mtt,mtt1} where order $\tl/J^2$ terms were computed to three
loops ($\tl^3$). We need to combine   the second order
perturbation theory correction for the quartic Hamiltonian
\rf{quartic} with the first order perturbation theory correction
 for the sixth
order Hamiltonian in \rf{six}. The regularization issues were
discussed in detail in \ci{mtt}: to match string/gauge results  we
 should use the normal-ordered form  of the Hamiltonians and 
apply $\zeta$-function  regularization for intermediate-state
sums. We shall also need to add a local higher-derivative
``counterterm'' which (on gauge side) is a lattice correction to
the continuum limit of the LL action (see \ci{mtt} and below). We
present details of this computation in Appendix B.

We find the following $\tl^4/J^2$ corrections to BMN energies
\begin{eqnarray}
E^{(2)}=&& \frac{\tl^4}{J^2}\bigg[\bigg(\frac{429}{16}+\frac{3
\a}{2}-512 \ \a_{1}-864 \ \a_{2}+1664 \ \a_{3}+576 \ \a_{4}-256 \
\a_{5}\nonumber\\
&& \ \ \ \ \ \ + \ 64 \ \a_{6}-\frac{3
\b}{4}\bigg)\sum_{i=1}^{M}n_{i}^2
\bigg(\sum_{j=1}^{M} n_{j}^3\bigg)^2\nonumber\\
&& \ \ +\bigg(-\frac{1087}{32}+\frac{5 \a}{4}+128 \ \a_{1}+1632
\ \a_{2}-2496 \ \a_{3}-1312 \ \a_{4}-128 \ \a_{5}\nonumber\\
&& \ \ \ \ \ \ - \ 64 \ \a_{6}+\frac{7
\b}{8}\bigg)\sum_{i=1}^{M}n_{i}^4 \bigg(\sum_{j=1}^{M}
n_{j}^2\bigg)^2\nonumber\\
&& \ \ +\bigg(\frac{1}{16}-704 \ \a_{1}-1472
\ \a_{2}+2496 \ \a_{3}+896 \ \a_{4}+128 \ \a_{5}\nonumber\\
&& \ \ \ \ \ \ + \ 64 \ \a_{6}-\frac{3
\b}{2}\bigg)\bigg(\sum_{i=1}^{M}n_{i}^4\bigg)^2\nonumber\\
&& \ \ + \ \bigg(-\frac{477}{32}+\frac{13 \a}{4}+64 \ \a_{1}+416 \
\a_{2}-704 \ \a_{3}-320 \ \a_{4}-768 \ \a_{5}+\frac{3
\b}{8}\bigg)\sum_{i=1}^M
n_{i}^8\nonumber\\
&& \ \  + \ \bigg(\frac{189}{8}-4 \a+256 \ \a_{1}-1984 \
\a_{2}+2688
\ \a_{3}+1728 \ \a_{4}+512 \ \a_{5}\nonumber\\
&& \ \ \ \ \ \ + \ 64 \ \a_{6}-\b \bigg) \sum_{i=1}^M n_{i}^6
\sum_{j=1}^M n_{j}^2\nonumber\\
&& \ \ + \ \bigg(-\frac{23}{8}-2 \a+768 \ \a_{1}+2240
\ \a_{2}-3584 \ \a_{3}-1536 \ \a_{4}+512 \ \a_{5}\nonumber\\
&& \ \ \ \ \ \ - \ 128 \ \a_{6}+2 \b\bigg)\sum_{i=1}^M n_{i}^5
\sum_{j=1}^M n_{j}^3\nonumber\\
&& \ \ + \ \frac{1}{4}\bigg(-\frac{21}{4}+4 \a-512(\a_{1}-\
\a_{2}+\ \a_{3}+\ \a_{4})\bigg)\sum_{i\neq j}\frac{n_{i}^5
n_{j}^5}{(n_{i}-n_{j})^2}\nonumber\\
&& \ \ \ \ \ \ + \ \frac{5\pi^2}{96}\sum_{i=1}^{M}n_{i}^{10}\bigg]
\ , \label{final}
\end{eqnarray}
where, as we mentioned in section 2, we added as in
\cite{mtt,mtt1} the expansion to four loops of the second term in
kinetic term (\ref{kinetic}).

\section{Comparison with Bethe ansatz computation and
fixing the LL Lagrangians}

In this section we are matching the $\tl^4/J^2$ corrections
obtained above from the quantum LL and the corresponding
corrections obtained from Bethe ansatz. We will thus fix the
coefficients at four loops for both string and gauge LL
Lagrangians.

In \cite{mtt1} the gauge Bethe ansatz (BDS) and string Bethe
ansatz (AFS) were used to find the $1/J^2$ corrections to all
orders in $\tl$. We present here the results to four loops. In the
case of the gauge theory we find
 \begin{eqnarray}\label{E2final1}
E^{(2)}_g&=&\frac{\tl}{J^2}\left[-\frac{\pi^2}{6}\sum_{i=1}^{M}n_{i}^4
+2\sum_{i=1}^{M}n_{i}^2 -\sum_{j\ne i}^M\frac{2n_i^2n_j^2}{ (n_i-n_j)^2}\right]\nonumber\\
&+&\frac{\tl^2}{J^2}\left[\frac{\pi^2}{12}\sum_{i=1}^{M}n_{i}^6
-\frac{13}{2}\sum_{i=1}^{M}n_{i}^4+\frac{3}{2}\bigg(\sum_{i=1}^{M}n_{i}^2\bigg)^2+\sum_{i\ne j}^M\frac{n_i^3n_j^3}{(n_i-n_j)^2}\right]\nonumber\\
&+&\frac{\tl^3}{J^2}\Bigg[-\frac{\pi^2}{16}\sum_{i=1}^{M}n_{i}^8+\frac{49}{4}\sum_{i=1}^{M}n_{i}^6-\frac{9}{4}\sum_{i=1}^{M}n_{i}^4
\sum_{j=1}^{M}n_{j}^2-\frac{9}{2}\bigg(\sum_{i=1}^{M}n_{i}^3\bigg)^2\nonumber\\
&& \ \ \ \ \ \ \  -\frac{1}{4}\bigg(\sum_{i=1}^{M}n_{i}^2\bigg)^3
-\frac{3}{4}\sum_{i\ne j}^M\frac{n_i^4n_j^4}{(n_i-n_j)^2}\Bigg]\nonumber\\
&+&\frac{\tl^4}{J^2}\Bigg[\frac{5\pi^2}{96}\sum_{i=1}^{M}n_{i}^{10}-\frac{305}{16}\sum_{i=1}^{M}n_{i}^8+\frac{15}{8}\sum_{i=1}^{M}n_{i}^6
\sum_{j=1}^{M}n_{j}^2+ \frac{41}{4}\sum_{i=1}^{M}n_{i}^5
\sum_{j=1}^{M}n_{j}^3\nonumber\\
&& \ \ \ \ \ \ \ +
\frac{25}{16}\left(\sum_{i=1}^{M}n_{i}^4\right)^2-
\frac{7}{8}\bigg(\sum_{i=1}^{M}n_{i}^3\bigg)^2\sum_{j=1}^{M}n_{j}^2+\frac{5}{8}\sum_{i=1}^{M}n_{i}^4\bigg(\sum_{i=1}^{M}n_{i}^2\bigg)^2\nonumber\\
&& \ \ \ \ \ \ \ +\frac{5}{8}\sum_{i\ne
j}^M\frac{n_i^5n_j^5}{(n_i-n_j)^2}\Bigg] +{\rm
O}(\frac{\tl^5}{J^2}) \ .
\end{eqnarray}
For the string case, the result is
\begin{eqnarray}\label{E2final2}
E^{(2)}_s&=&\frac{\tl}{J^2}\left[-\frac{\pi^2}{6}\sum_{i=1}^{M}n_{i}^4
+2\sum_{i=1}^{M}n_{i}^2 -\sum_{j\ne i}^M\frac{2n_i^2n_j^2}{ (n_i-n_j)^2}\right]\nonumber\\
&+&\frac{\tl^2}{J^2}\left[\frac{\pi^2}{12}\sum_{i=1}^{M}n_{i}^6
-\frac{13}{2}\sum_{i=1}^{M}n_{i}^4+\frac{3}{2}\bigg(\sum_{i=1}^{M}n_{i}^2\bigg)^2+\sum_{i\ne j}^M\frac{n_i^3n_j^3}{(n_i-n_j)^2}\right]\nonumber\\
&+&\frac{\tl^3}{J^2}\Bigg[-\frac{\pi^2}{16}\sum_{i=1}^{M}n_{i}^8+\frac{49}{4}\sum_{i=1}^{M}n_{i}^6-\frac{31}{8}\sum_{i=1}^{M}n_{i}^4
\sum_{j=1}^{M}n_{j}^2-3\bigg(\sum_{i=1}^{M}n_{i}^3\bigg)^2\nonumber\\
&& \ \ \ \ \ \ \  +\frac{1}{8}\bigg(\sum_{i=1}^{M}n_{i}^2\bigg)^3
-\frac{3}{4}\sum_{i\ne j}^M\frac{n_i^4n_j^4}{(n_i-n_j)^2}\Bigg]\nonumber\\
&+&\frac{\tl^4}{J^2}\Bigg[\frac{5\pi^2}{96}\sum_{i=1}^{M}n_{i}^{10}-\frac{305}{16}\sum_{i=1}^{M}n_{i}^8+\frac{75}{16}\sum_{i=1}^{M}n_{i}^6
\sum_{j=1}^{M}n_{j}^2+ \frac{51}{8}\sum_{i=1}^{M}n_{i}^5
\sum_{j=1}^{M}n_{j}^3\nonumber\\
&& \ \ \ \ \ \ \ +
\frac{23}{8}\left(\sum_{i=1}^{M}n_{i}^4\right)^2-
\frac{7}{16}\bigg(\sum_{i=1}^{M}n_{i}^3\bigg)^2\sum_{j=1}^{M}n_{j}^2-\frac{5}{16}\sum_{i=1}^{M}n_{i}^4\bigg(\sum_{i=1}^{M}n_{i}^2\bigg)^2\nonumber\\
&& \ \ \ \ \ \ \ +\frac{5}{8}\sum_{i\ne
j}^M\frac{n_i^5n_j^5}{(n_i-n_j)^2}\Bigg] +{\rm
O}(\frac{\tl^5}{J^2}) \ .
\end{eqnarray}

Comparing (\ref{final}) with the gauge part (\ref{E2final1})
except for the pole type term we obtain six equations with six
unknowns, which solved give the gauge LL coefficients
\begin{equation}
\a_{1}^{(g)}=0 \ , \quad
\a_{2}^{(g)}=-\frac{64}{256}=-\frac{1}{4}\ , \quad
\a_{3}^{(g)}=-\frac{221}{2048}\ , \quad
\a_{4}^{(g)}=-\frac{175}{1024}
\end{equation}
\begin{equation}
\a_{5}^{(g)}=\frac{111}{4096} \ , \quad
\a_{6}^{(g)}=\frac{141}{256} \ .
\end{equation}
We see that the $\tl^4/J^2$ corrections contain enough details to
fix almost completely the gauge LL Lagrangian except for the
coefficient $a_{7}^{(g)}$. As a consistency check we see that
these match the coefficients and equations between coefficients
obtained from comparing $\tl^4/J$, and the energy of circular
string (\ref{g1}), (\ref{g2}), (\ref{g3}). The conclusion is that
besides finding the full gauge LL Lagrangian at four loops, we
also find perfect agreement for the $\tl^4/J$ and $\tl^4/J^2$
corrections as obtained from quantum gauge LL and gauge Bethe
ansatz.

Similarly, for the string part comparing (\ref{final}) with
(\ref{E2final2}) we find the coefficients
\begin{equation}
\a_{1}^{(s)}=0 \ , \quad \a_{2}^{(s)}=-\frac{67}{256} \, \quad
\a_{3}^{(s)}=-\frac{237}{2048}\ , \quad
\a_{4}^{(s)}=-\frac{179}{1024} \ ,
\end{equation}
\begin{equation}
\a_{5}^{(s)}=\frac{119}{4096} \ , \quad
\a_{6}^{(s)}=\frac{649}{1024} \ .
\end{equation}
These are again consistent with our findings in (\ref{s1}),
(\ref{s2}), (\ref{s3}). We find again the string LL Lagrangian at
four loops and also perfect agreement for the $\tl^4/J$ and
$\tl^4/J^2$ corrections as obtained from quantum string LL and
string Bethe ansatz.

Let us now look at the pole type term. This is the same as
obtained from both string and gauge Bethe ansatze. Matching the
pole term we obtain
\begin{equation}
\frac{1}{4}\bigg(-\frac{21}{4}+4
\a-512(\a_{1}-\a_{2}+\a_{3}+\a_{4})\bigg)=\frac{5}{8} \ ,
\end{equation}
which gives
\begin{equation}
\a_{1}-\a_{2}+\a_{3}+\a_{4}=-\frac{59}{2048} \ ,
\end{equation}
which is consistent with the equations (\ref{s1}), (\ref{g1}).
Note that the pole type term depends only on the coefficient $ \a$
and the combination $\a_{1}-\a_{2}+\a_{3}+\a_{4}$, which are the
same for both gauge and string LL.

As expected we found that the disagreement between string LL and
gauge LL actions continues also at four loops. The difference
between the two LL Lagrangians or Hamiltonians is
\begin{eqnarray}\la{uup}
L_{s}-L_{g}&=& - ( H_s - H_g) = \frac{\tilde{\lambda}^3}
{64}\left[(\partial_{1}\vn\cdot\partial_{1}^2
\vn)^2-\frac{1}{16}(\partial_{1}\vn\cdot\partial_1\vn)^3\right]\nonumber\\
&-&\frac{\tl^4}{256}\bigg[3 \ (\partial_{1}
\vec{n})^2(\partial_{1}^2 \vec{n} \partial_{1}^4 \vec{n}) +2 \
(\partial_{1}\vec{n}\partial_{1}^5\vec{n})(\partial_{1}\vec{n})^2+
\ (\partial_{1}^3
\vec{n})^2(\partial_{1}\vec{n})^2\nonumber\\
&-& \ \frac{1}{2} \ (\partial_{1}\vec{n})^4(\partial_{1}^2
\vec{n})^2- \ \frac{85}{4} \ (\partial_{1}\vec{n}\partial_{1}^2
\vec{n})^2(\partial_{1}\vec{n})^2+\frac{7}{16} \
(\partial_{1}\vec{n})^8\bigg]+ O(\tilde{\lambda}^5)
\end{eqnarray}
We observe that both at three-loops and four-loops only one of the
terms has the same coefficients in both gauge and string LL
Lagrangians, i.e. $\a_s=\a_g$ at three-loops and
$\a_1^{(s)}=\a_{1}^{(g)}$ at four-loops. The disagreement between
the two LL Lagrangians is not unexpected\footnote{It is rather
remarkable that the gauge and string LL actions are the same at
one and two-loops.} and can be explained by the order of limits
taken on the two sides \cite{ss,bds}. On string theory side one
first takes $J$ large with $\tl$=fixed to suppress quantum
corrections and then expands in small $\tl$, while on gauge theory
side $\lambda$ small is taken first, and then large $J$ expansion
to isolate contributions depending on $\tl$.

\bigskip
\bigskip

\section*{Acknowledgments }

We are grateful to A. Tseytlin for discussions and comments on the
manuscript. We also thank J. Minahan for discussions and sharing
with us unpublished Bethe ansatz results. The work of A.T.
  was supported  by the DOE grant DE-FG02-91ER40690.

\renewcommand{\theequation}{A.\arabic{equation}}
 \setcounter{equation}{0}
\setcounter{section}{1} \setcounter{subsection}{0}

  \section*{Appendix A:
  Fixing coefficients in  4-loop gauge and string LL actions:
  circular string example}

In this section we recall the  details of the rational circular
string solution in the $SU(2)$ sector \cite{art} and compute its
classical energy to order $\tl^4$. The solution is given by
\begin{equation}
{\rm X}_{r}=b_{r}e^{i(w_{r}\tau+m_{r}\sigma)}\ ,\ \ \ \ \ \  \quad
r=1,2
\end{equation}
where $\XX_{r}^2 =1 $ are  $S^3$ coordinates and
\begin{equation}
b_{1}^2+b_{2}^2=1\ , \quad \quad w_{r}=\sqrt{m_{r}^2+\nu^2} \ , \
\ \ \ J_r= \sqrt \l \mathcal{J}_{r} = \sqrt \l b_{r}^2w_{r}\ . \ee
$\nu$ is a parameter to be determined from the conformal gauge
constraints
\begin{equation}
\mathcal{E}^2=2 ( w_{1}\mathcal{J}_{1} +w_{2}\mathcal{J}_{2})
  -\nu^2\ , \ \ \ \ \ \ \ \quad \quad
m_{1}\mathcal{J}_{1} + m_2 \mathcal{J}_{2} =0 \ ,
\end{equation}
where the energy is $E=\sqrt{\lambda}\mathcal{E}$.
 Introducing the notation
$$m\equiv m_{1}\ , \ \ \ \ \ \
n\equiv m_{1}-m_{2}\ , \ \ \ \ \
\mathcal{J}=\mathcal{J}_{1}+\mathcal{J}_{2}\  ,
$$ one can solve one of the
constraints for $\nu$ at large $\mathcal{J}$  or small $\tl= { 1
\ov \mathcal{J}^2}$ to  obtain
\begin{eqnarray}
\nu^2&=&\mathcal{J}^2 + m(m -n) - \frac{3 m(m - n)(2 m -
n)^2}{4\mathcal{J}^2} + \frac{5 m(m - n)(2 m - n)^4}{8
\mathcal{J}^4}\nonumber\\
&-& \frac{7}{64 \mathcal{J}^6}m(n-2m)^4(24 m^3-48 nm^2+29 n^2 m-5
n^3)\nonumber\\
&+&\frac{9}{128 \mathcal{J}^8}m(n-2m)^6 (44 m^3-88 n m^2 +51 n^2
m-7 n^3)+O({ 1 \ov \mathcal{J}^{10}}).
\end{eqnarray}
Then the string energy to $\tl^4$ order is found to be
\begin{eqnarray}
E_s &=&  J \bigg[ 1 +  \frac{1}{2} \tl  m (n-m)
                          - \frac{1}{8} \tl^2  m (n-m) (n^2 - 3 m n  + 3
                          m^2)\nonumber\\
     & +& \ \frac{1}{16} \tl^3 m (n-m) (n^4 - 7 m n^3  + 20 m^2 n^2 - 26 m^3 n +
     13m^4)\nonumber\\
     &+& \frac{1}{128}\tl^4 m(m-n)(323 m^6 -969 nm^5+1207 n^2
     m^4-799 n^3 m^3+297 n^4 m^2\nonumber\\
     &-&59 n^5 m+5 n^6)
                         +O(\tilde{\lambda}^5)\bigg] \ .
             \la{sti}
\end{eqnarray}

\bigskip

Starting  now with the LL Lagrangian (\ref{4loop}) let us  find
the energy for the corresponding
 solution with $J_{1} \ne
J_{2}$  which is given to leading order by $\vn= (n_1,n_2,n_3)$
where \cite{krt}
\begin{equation}
n_{1}=2\sqrt{\frac{m}{n}\left(1-\frac{m}{n}\right)}\cos
n\sigma+O(\tilde{\lambda}), \quad \quad
n_{2}=2\sqrt{\frac{m}{n}\left(1-\frac{m}{n}\right)}\sin
n\sigma+O(\tilde{\lambda}),
\end{equation}
\begin{equation}
 n_{3}=1-\frac{2m}{n}+O(\tilde{\lambda}) \ .
\end{equation}
This solution can also be found  by expanding the full string
solution at large $\mathcal{J}$.\foot{The unit vector $\vec{n}$
can be written as $\vec{n}=(\sin 2\psi \cos 2\varphi,\sin 2\psi
\sin 2\varphi, \cos 2\psi)$. In terms of global angular
coordinates of $S^5$ with the metric $ds^2=dt^2+d\gamma^2+\cos^{2}
\gamma\ d\varphi_{3}^2+\sin^{2} \gamma\ (d\psi^2+\cos^{2}\psi\
d\varphi_{1}^2+\sin^{2}\psi\ d\varphi_{2}^2) $ we have
$\varphi=\frac{\varphi_{1}-\varphi_{2}}{2}$. Note also that the
cartesian coordinates are ${\rm X}_{1}=\cos \psi\ e^{i
\varphi_{1}}$, ${\rm X}_{2}=\sin \psi\ e^{i \varphi_{2}}.$}
Plugging this solution into the  Hamiltonian in (\ref{4loop}) we
find for its    LL  energy
\begin{eqnarray}\label{ELL}
E_{_{LL}}&=&
J\bigg[1+\frac{\tilde{\lambda}}{2}m(n-m)-\frac{\tilde{\lambda}^2}{8}m(n-m)(3m^2+n^2-3
m n)\nonumber\\
&+& \ \frac{\tilde{\lambda}^3}{16}m(n-m)\left[n^4-7 \ n^2 m (n-m)
 + 13  \ m^2(n-m)^2  \right]\nonumber\\
 &+&\frac{\tl^4}{128} m(m - n) \ [\- 32768 \ \a_{7} \ m^6 + 98304 \ \a_{7} \ m^5 n
 -
        98304 \ \a_{7} \ m^4 n^2 + 8192 \ \a_{5} \ m^4 n^2 \nonumber\\
        &+ &32768 \ \a_{7} \ m^3 n^3 - 16384 \ \a_{5} \ m^3 n^3 -
        2048 \ \a_{1} \ m^2 n^4  - 2048 \ \a_{3} \ m^2 n^4 \nonumber\\
        &+& 8192 \ \a_{5} \ m^2 n^4 - 2048 \ \a_{4} \ m^2 n^4+ 2048 \ \a_{2} \ m^2 n^4
        + \ 2048 \ \a_{1} \ m n^5
        \nonumber\\
        & +& 2048 \ \a_{3} \ m n^5
         +2048 \ \a_{4} \ m n^5- 2048 \ \a_{2} \ m n^5 +5 \ n^6]+
         O(\tilde{\lambda}^5)\bigg] \ .
\end{eqnarray}
Comparing (\ref{sti}) with (\ref{ELL}) we obtain the coefficients
\begin{equation}
\a_{5}^{(s)}=\frac{119}{4096}, \quad \quad
\a_{7}^{(s)}=-\frac{323}{32768}, \quad \quad
\a_{1}^{(s)}-\a_{2}^{(s)}+\a_{3}^{(s)}+\a_{4}^{(s)}=-\frac{59}{2048}.
\end{equation}

\bigskip

We want now to compare the four-loop LL energy for the circular
solution with the energy of the corresponding state on the gauge
theory side obtained from the Bethe ansatz. One can compute the
latter as in \cite{m}, but now using BDS Bethe ansatz, and the
result is\footnote{We are grateful to J. Minahan for sharing these
unpublished results with us. Let us also mention that the
computation of energy for this solution has also been computed
using the string Bethe ansatz and the results match (\ref{sti}),
as expected.}
\begin{eqnarray}
E_g &=&  J \bigg[ 1 +  \frac{1}{2} \tl  m (n-m)
                          - \frac{1}{8} \tl^2  m (n-m) (n^2 - 3 m n  + 3
                          m^2)\nonumber\\
     & +&\ \frac{1}{16} \tl^3 m (n-m) (n^2-3 \ mn+3 \ m^2)(n-2 \ m)^2\nonumber\\
     &-& \frac{1}{128}\tl^4 m(n-m)(267 \ m^6 -801 \ nm^5+1023 \
     n^2m^4-711 \ n^3 m^3\nonumber\\
     &+&281 \ n^4 m^2-59 \ n^5 m+5 \
     n^6)          +O(\tilde{\lambda}^5)\bigg] \ .
             \la{gauge}
\end{eqnarray}
Comparing (\ref{ELL}) with (\ref{gauge}) we obtain the following
coefficients and equations between them
\begin{equation}
\a_{5}^{(g)}=\frac{111}{4096}, \quad \quad
\a_{7}^{(g)}=-\frac{267}{32768}, \quad \quad
\a_{1}^{(g)}-\a_{2}^{(g)}+\a_{3}^{(g)}+\a_{4}^{(g)}=-\frac{59}{2048}.
\end{equation}

\renewcommand{\theequation}{B.\arabic{equation}}
 \setcounter{equation}{0}
\setcounter{section}{1} \setcounter{subsection}{0}
 \section*{Appendix B: More on the $1/J^2$ corrections
 to BMN energy from quantum LL}

In this appendix we present some details of the computation of
$\tl^4/J^2$ corrections to the BMN energies. We start with the
second-order perturbation (``exchange'')   contribution. Starting
with the  quartic Hamiltonian \rf{hi} we  need to compute
\begin{equation}\la{secc}
 \langle  M| (\bar H_4)^{(2)} |M\rangle=
  \sum_{M\neq
M'}\frac{\langle  M|\bar H_4 |M'\rangle \langle  M'| \bar H_4
|M\rangle}{E_M-E_{M'}}  \  ,
\end{equation}
where $|M'\rangle $ is any possible intermediate state, and
 $|M\rangle=a^\dagger _{n_{1}} ... a^\dagger_{n_{M}}|0\rangle$.
 Since $\bar H_4$ in \rf{hi} contains only terms of  the
form $(a^{\dagger})^2a^2$, the only non-trivial intermediate
states can be the $M'=M$ -particle states 
 of the form $ a^\dagger
_{n'_{1}} ... a^\dagger_{n'_{M}}|0\rangle  $.
 Then in order for
the matrix element $\langle  0 | a_{n_{1}} ... a_{n_{M}}
 |\bar H_4 | a^\dagger _{n'_{1}} ...
a^\dagger_{n'_{M}}|0\rangle $ to be non-zero, there should be a
$j$ and $k$ such that   $n'_j=n_j+q$ and $n'_k= n_k-q$, with all
other $n_i'=n_i$, $i\ne j,k$.  In order for $|M\rangle$ to be
distinct from $|M'\rangle$, we require that $0\ne q\ne n_k-n_j$.
With these conditions, we then find that if $n_k\ne n_j$
\begin{eqnarray} \la{uio}
\HH_{4}^{(1)}\equiv &&\langle  M|\bar H_4 |M'\rangle
=\frac{1}{J}\bigg\{  {2}\sqrt{1+\tilde{\lambda}q^2}
+2\sqrt{1+\tilde{\lambda}(n_{k}-n_{j}-q)^2} \nonumber\\ &&  - \
\sqrt{1+\tilde{\lambda}n_{k}^2}-
\sqrt{1+\tilde{\lambda}n_{j}^2}-\sqrt{1+
\tilde{\lambda}(q+n_{j})^2} - \sqrt{1+\tilde{\lambda}(n_{k}-q)^2}
\nonumber\\
&&    +  \ \frac{3\tilde{\lambda}^2}{2}n_{k}n_{j}(q+n_{j})
(n_{k}-q)\bigg[1+\frac{\tilde{\lambda}}{12}\bigg(2\a
(n_{j}+n_{k})^2   + \ \b[(n_{k}-n_{j})^2-2q
(n_{k}-n_{j}-q)] \bigg)\bigg]\nonumber\\
&&  - 8 \ \tl^4 \bigg[8 \a_1 n_{j}^2 n_{k}^2
(n_{j}+q)^2(n_{k}-q)^2+\a_2n_{j}n_{k}(n_{j}+q)(n_{k}-q)[2(q+n_{j})^2(n_{k}-q)^2\nonumber\\
&&  +
2n_{j}^2n_{k}^2+(n_{j}^2+n_{k}^2)((n_{j}+q)^2+(n_{k}-q)^2)]-2 \a_3
n_{j}n_{k}^2(q+n_{j})(n_{k}-q)[n_{k}^2(n_{j}+n_{k})\nonumber\\
&&  +(q+n_{j})^3+(n_{k}-q)^3]+2\a_4
n_{j}n_{k}(q+n_{j})(n_{k}-q)[2n_{k}^4+(q+n_{j})^4+(n_{k}-q)^4]\nonumber\\
&&  +4 \a_5
n_{j}n_{k}^3(n_{j}+q)(n_{k}-q)[(n_{j}+q)^2+(n_{k}-q)^2]\bigg]\bigg\},\
\end{eqnarray}
 where $n_j+q$ and  $n_k-q$ are not equal to
one of the other $n_l$'s. The energy difference in \rf{secc} is
\begin{equation}\la{kk}
W_{1}\equiv
E_M-E_{M'}=\sqrt{1+\tilde{\lambda}n_{j}^2}+\sqrt{1+\tilde{\lambda}n_{k}^2}
-\sqrt{1+\tilde{\lambda}(n_{j}+q)^2}-\sqrt{1
+\tilde{\lambda}(n_{k}-q)^2}\ .
\end{equation}
 If $n_j+q=n_l$,
and so $|M'\rangle$ has two impurities with the same momenta, then
the matrix element is
\begin{eqnarray}\la{oop}
\HH_{4}^{(2)}\equiv &&\langle  M|\bar H_4 |M'\rangle
=\frac{\sqrt{2}}{J}\HH_{4}^{(1)}|_ {\ q=n_{l}-n_{j}}
\end{eqnarray}
and the energy difference is
\begin{equation}
W_{2}\equiv
E_M-E_{M'}=\sqrt{1+\tilde{\lambda}n_{j}^2}+\sqrt{1+\tilde{\lambda}n_{k}^2}
-\sqrt{1+\tilde{\lambda}n_{l}^2}-\sqrt{1+
\tilde{\lambda}(n_{j}+n_{k}-n_{l})^2} .
\end{equation}
Then  the ``exchange'' contribution is given by
\begin{equation}\label{H42sum}
\langle  M| (\bar H_4)^{(2)} |M\rangle
=\frac{1}{4}\sum^M_{j,k\atop j \ne k}\bigg[
\sum_{q=-\infty\atop0\ne q\ne n_k-n_j}^\infty
\frac{(\HH_{4}^{(1)})^2}{W_{1}}+\sum^M_{l\ne j\atop l\ne
k}\frac{(\HH_{4}^{(2)})^2}{W_{2}} \  \bigg] \ .
\end{equation}
It was seen in \cite{mtt1} that if one looks at this expression as
a string-theory expression (i.e. non-perturbative in $\tl$, with
sums done before the expansion in $\tl$), the sum over the
``virtual'' momentum $q$ produces a contribution which is
non-analytic in $\tl$. This phenomenon, which was absent at order
$1/J$, was first observed \ci{bt}. As we already pointed out, it
was shown in \cite{mtt1} that the correct coefficients of the
non-analytic terms cannot be determined correctly in the
Landau-Lifshitz approach as other modes outside the $SU(2)$ sector
contribute to non-analytic terms. In this paper we focus only on
the analytic terms obtained by expanding in $\tl$ before doing the
sum over $q$. Our aim here is to extend  the computation of
\ci{mtt,mtt1} to the order $\tilde{\lambda}^4/J^2$ and to show
that the results match
 the gauge and string  Bethe ansatz results. Computing the sums
 in (\ref{H42sum})
  we obtain the ``exchange''   contribution to the
 $\tl^4/J^2$ correction to the BMN energies
\begin{eqnarray}
&& \frac{\tl^4}{J^2}\bigg[\bigg(\frac{127}{8}+\frac{3 \a}{2}-384 \
\a_{1}-96 \ \a_{2}-128 \ \a_{3}+544 \ \a_{4}-32 \
\a_{5}\nonumber\\
&& \ \ \ \ \ \ -\frac{3 \b}{4}\bigg)\sum_{i=1}^{M}n_{i}^2
\bigg(\sum_{j=1}^{M} n_{j}^3\bigg)^2\nonumber\\
&& \ \ +\bigg(-\frac{599}{64}+\frac{5 \a}{4}+64 \ \a_{1}-144
\ \a_{2}+288 \ \a_{3}-576 \ \a_{4}-160 \ \a_{5}\nonumber\\
&& \ \ \ \ \ \ +\frac{7 \b}{8}\bigg)\sum_{i=1}^{M}n_{i}^4
\bigg(\sum_{j=1}^{M}
n_{j}^2\bigg)^2\nonumber\\
&& \ \ +\bigg(\frac{525 M}{128}-\frac{349}{16}-384 \ \a_{1}+32
\ \a_{2}-416 \ \a_{3}+896 \ \a_{4}+64 \ \a_{5}\nonumber\\
&& \ \ \ \ \ \ -\frac{3
\b}{2}\bigg)\bigg(\sum_{i=1}^{M}n_{i}^4\bigg)^2\nonumber\\
&& \ \ + \ \bigg(\frac{5M^2}{128}-\frac{477}{32}+\frac{13
\a}{4}-320 \ \a_{1}-512 \ \a_{2}+800 \ \a_{3}-1088 \ \a_{4}-704 \
\a_{5}\nonumber\\
&& \ \ \ \ + \frac{3 \b}{8}-5M\bigg)\sum_{i=1}^M
n_{i}^8\nonumber\\
&& \ \  + \ \bigg(\frac{105 M}{32}-\frac{217}{16}-4 \a+256 \
\a_{1}+352\ \a_{2}-544
\ \a_{3}+640 \ \a_{4}+448 \ \a_{5}\nonumber\\
&& \ \ \ \ \ \ -\b \bigg) \sum_{i=1}^M n_{i}^6
\sum_{j=1}^M n_{j}^2\nonumber\\
&& \ \ + \ \bigg(\frac{327}{8}-2 \a+768 \ \a_{1}+352
\ \a_{2}-448 \ \a_{4}+384 \ \a_{5}+2 \b \nonumber\\
&& \ \ -\frac{35M}{16}\bigg)\sum_{i=1}^M n_{i}^5
\sum_{j=1}^M n_{j}^3\nonumber\\
&& \ \ + \ \frac{1}{4}\bigg(-\frac{21}{4}+4 \a-512(\a_{1}-\
\a_{2}+\ \a_{3}+\ \a_{4})\bigg)\sum_{i\neq j}\frac{n_{i}^5
n_{j}^5}{(n_{i}-n_{j})^2}\nonumber\\
&& \ \ \ \ \ \ + \ \frac{5\pi^2}{96}\sum_{i=1}^{M}n_{i}^{10}\bigg]
\ , \label{secondorder}
\end{eqnarray}

\bigskip

\bigskip

Let us now compute the  sixth-order ``contact'' contribution
coming from the expectation value of the sixth order term in the
LL Hamiltonian (\ref{six}). The normal ordered form for this term
can be written as\foot{As in \cite{mtt1} the only
 regularization we use here is the assumption that the
sixth order term in the Hamiltonian is normal ordered.}
\begin{equation}
\bar{H}_{6} = \frac{1}{J^2}\sum_{n,m,k}\  h_{nmk} \
a_{n}^{\dagger}a_{m}^{\dagger}a_{k}^{\dagger}a_{n}a_{m}a_{k}\ ,
\la{hhh}
\end{equation}
where for simplicity we do not write down the complicated form of
$h_{nmk}$. After a lengthy but straightforward computation we
obtain the expectation value of $\bar{H}_{6}$
\begin{eqnarray}
&& \frac{\tl^4}{J^2}\bigg[\bigg(\frac{175}{16}-128 \ \a_{1}+112 \
\a_{2}-736 \ \a_{3}+1120 \ \a_{4}+608 \
\a_{5}\nonumber\\
&& \ \ \ \ \ \ - \ 256 \ \a_{6}+64 \
a_{7}\bigg)\sum_{i=1}^{M}n_{i}^2
\bigg(\sum_{j=1}^{M} n_{j}^3\bigg)^2\nonumber\\
&& \ \ +\bigg(-\frac{1575}{64}+64 \ \a_{1}-576
\ \a_{2}+1344 \ \a_{3}-1920 \ \a_{4}-1152 \ \a_{5}\nonumber\\
&& \ \ \ \ \ \ - \ 128 \ \a_{6}- \ 64 \
\a_{7}\bigg)\sum_{i=1}^{M}n_{i}^4 \bigg(\sum_{j=1}^{M}
n_{j}^2\bigg)^2\nonumber\\
&& \ \ +\bigg(\frac{175}{8}-320 \ \a_{1}-1056
\ \a_{3}+1600 \ \a_{4}+832 \ \a_{5}+128 \ \a_{6}\nonumber\\
&& \ \ \ \ \ \ + \ 64 \ \a_{7}-\frac{525 M}{128}
+\ 48 M \a_{2}\bigg)\bigg(\sum_{i=1}^{M}n_{i}^4\bigg)^2\nonumber\\
&& \ \ + \ \bigg(-\frac{5 M^2}{128}+384 \ \a_{1}+384 \ \a_{2}-384
\ \a_{3}+384 \ \a_{4}+384 \ \a_{5}-768 \ \a_6 \nonumber\\
&& \ \ \ \ \ \ + \ 5M-64 M \ \a_2 \bigg)\sum_{i=1}^M
n_{i}^8\nonumber\\
&& \ \  + \ \bigg(\frac{595}{16}+896 \ \a_{2}-1440 \ \a_{3}+2048
\ \a_{4}+1280 \ \a_{5}+512 \ \a_{6}\nonumber\\
&& \ \ \ \ \ \ + \ 64 \ \a_{7}-\frac{105 M}{32}+16 M \a_2 \bigg)
\sum_{i=1}^M n_{i}^6
\sum_{j=1}^M n_{j}^2\nonumber\\
&& \ \ + \ \bigg(-\frac{175}{4}-800
\ \a_{2}+2240 \ \a_{3}-3136 \ \a_{4}-1920 \ \a_{5}\nonumber\\
&& \ \ \ \ \ \ + \ 512 \ \a_{6}- \ 128 \a_7+\frac{35M}{16}
\bigg)\sum_{i=1}^M n_{i}^5 \sum_{j=1}^M n_{j}^3\bigg] \ ,
\label{sixorder}
\end{eqnarray}
Putting together (\ref{secondorder}) and (\ref{sixorder}) we see
that the explicit dependence on the number of impurities $M$
cancels, and we obtain the result (\ref{final}).

\end{document}